\documentclass[superscriptaddress,12pt]{revtex4}
\usepackage{amsfonts}
\usepackage{graphicx}
\usepackage{marvosym}
\begin{document}
\title{Adiabatic passage for three-dimensional
entanglement generation through quantum Zeno dynamics {\footnote{Yan Liang $\cdot$ Shi-Lei Su $\cdot$  Xin Ji ({\large \Letter})  $\cdot$ Shou Zhang\\
 Department of Physics, College of Science, Yanbian University,
\\ Yanji, Jilin 133002, People's Republic of China \\ Qi-Cheng Wu $\cdot$ Department of Physics, College of Science,  Fuzhou University,
\\ Fuzhou, Fujian 350108, People's Republic of China\\
 e-mail: jixin@ybu.edu.cn}}}
\author{\textbf{Yan Liang, Shi-Lei Su, Qi-Cheng  Wu, Xin Ji, Shou Zhang} }

\begin{abstract}
\noindent \textbf{{Abstract}}: We propose an adiabatic passage
approach to generate two atoms three-dimensional entanglement with
the help of quantum Zeno dynamics in a time-dependent interacting
field. The atoms are trapped in two spatially separated cavities
connected by a fiber, so that the individual addressing is
needless.~Because the scheme is based on the resonant interaction,
the time required to generate entanglement is greatly
shortened.~Since the fields remain in vacuum state and all the
atoms are in the ground states, the losses due to the excitation
of photons and the spontaneous transition of atoms are suppressed
efficiently compared with the dispersive protocols. Numerical
simulation results show that the scheme is robust against the
decoherences caused by the cavity decay and atomic spontaneous
emission. Additionally, the scheme can be generalized to generate
$N$-atom three-dimensional entanglement and high-dimensional
entanglement for two spatially separated atoms.
\end{abstract}

\maketitle Quantum entanglement, an interesting and attractive
phenomenon in quantum mechanics, plays a significant role not only
in testing quantum nonlocality, but also in a variety of quantum
information tasks \cite{AKE,CHB,CHBS,CHBG,KMH,MHVB,SBZGCG,GV},
such as quantum computing \cite{SHT1997,MI2000,CD2000},
teleportation \cite{CHBG}, cryptography \cite{AKE}, precision
measurements \cite{JWDD1996} and so on. Recently, high-dimensional
entanglement is becoming more and more important since they are
more secure than qubit systems, especially in the aspect of
quantum key distribution. Besides, it has been demonstrated that
violations of local realism by two entangled high-dimensional
systems are stronger than that by two-dimensional systems
\cite{DPMW2000}.~A lot of works have been done in generation of
high-dimensional entanglement. For example, Wu $et~al.$ proposed a
scheme for generating a multiparticle three-dimensional
entanglement by appropriately adiabatic evolutions
\cite{WCYC2013}. Li and Huang deterministically generated a
three-dimensional entanglement via quantum Zeno dynamics(QZD)
\cite{WG2011}. Chen $et~al.$ proposed a scheme to prepare
three-dimensional entanglement state between a single atom and a
Bose-Einstein condensate (BEC) via stimulated Raman adiabatic
passage (STIRAP) technique \cite{LP2012}. In experiment, two
schemes have been put forward to generate high-dimensional
entanglement by the means of the spatial modes of the
electromagnetic field carrying orbital angular momentum
\cite{AAGA2001,AGA2002}.

In order to realize the entanglement generation or population
transfer in a quantum system with time-dependent interacting
field, many schemes have been put forward. Such as $\pi$ pulses,
composite pulses, rapid  adiabatic passage(RAP), stimulated Raman
adiabatic passage , and their variants
\cite{KHB1998,PIMS2007,NVTB2001}. STIRAP is widely used in
time-dependent interacting field because of the robustness for
variations in the experimental parameters. But it usually requires
a relatively long interaction time, so that the decoherence would
destroy the intended dynamics, and finally lead to an error
result. Therefore, reducing the time of dynamics towards the
perfect final outcome is necessary and perhaps the most effective
method to essentially fight against the dissipation which comes
from noise or losses accumulated during the operational processes.
Rencently, various schemes have been explored theoretically and
experimentally to construct shortcuts for adiabatic passage
\cite{ARXD2012,XASA2010,KPYR2011,AC2013,MYLJ2014,YYQJ2014,AFTS2012,JXPP2011}.

 On the other hand, the quantum Zeno effect is an interesting
phenomenon in quantum mechanics. It stems from general features of
the Schr$\ddot{\rm{o}}$dinger equation that yield quadratic
behavior of the survival probability at short time
\cite{HMS1996,PPAL2001}. The quantum Zeno effect, which has been
tested in many experiments, is the inhibition of transitions
between quantum states by frequent measurements
\cite{WMDJ1990,PHTA1995}. Recent studies
\cite{PVGS2000,PS2002,PGS2009} show that a quantum Zeno evolution
will evolve away from its initial state, but it remains in the
¡°Zeno subspace¡± defined by the measurements
\cite{PVGS2000,PPAL2001} via frequently projecting onto a
multidimensional subspace. This is known as QZD. Suppose that a
dynamical evolution of a system can be governed by the Hamiltonian
$H_K=H_{\rm obs}+KH_{\rm meas}$, where $H_{\rm obs}$ is the
Hamiltonian of the investigated quantum system and the $H_{\rm
meas}$ is regarded as an additional interaction Hamiltonian
performing the measurement, while $K$ is a coupling constant. In
the limit $K\rightarrow \infty$, the system is governed by the
evolution operator $U(t)={\rm
exp}[-it\sum_{n}(K\lambda_nP_n+P_nH_{\rm obs}P_n)]$, which is an
important basis for our following work, with $P_n$ is the
eigenprojections of $H_{\rm meas}$ with eigenvalues
$\lambda_n$($H_{\rm meas} = \sum_{n}\lambda_nP_n$).

 In this paper, we present an effective scheme to construct an
adiabatic passage for three-dimensional entanglement generation
between atoms motivated by the space division of QZD. The atoms
are individually trapped in distant optical cavities connected by
a fiber. Compared with previous works, our scheme has the
following advantages: First, the two atoms three-dimensional
entanglement can be achieved in one step, which will effectively
reduce the complexity for implementing the scheme in experiment.
Second,~our scheme is based on the resonant interaction so the
evolution time is very short.~Third, the scheme is very robust
against the photons leakage and atoms decay since the system only
evolves in the null-excitation subspace.~Fourth, the scheme can be
expanded to generate $N$-atom three-dimensional entanglement and
high-dimensional entanglement.~This paper is structured as
follows: In Sec.~2, we construct the fundamental model and give
the effective dynamics to generate three-dimensional entanglement
of two spatially separated atoms. In Sec.~3, we analyze the
robustness of this scheme via numerical simulation. In Sec.~4, we
generalize this proposal to generate $N$-atom three-dimensional
entanglement.~Besides, in Sec.~5, we expand our scheme to generate
high-dimensional entanglement between two distant atoms. The
conclusion appears in Sec.~6.

\begin{figure}[htb]\centering
\includegraphics[width=7cm]{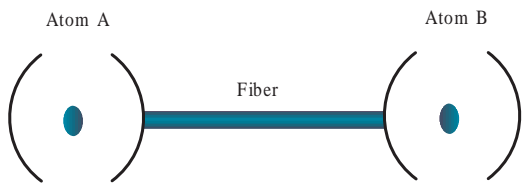}
\caption{The schematic setup for generating two atoms
three-dimensional entanglement. The two atoms are trapped in two
spatially separated optical cavities connected by a fiber.}
\end{figure}
\begin{figure}[htb]\centering
\includegraphics[width=10cm]{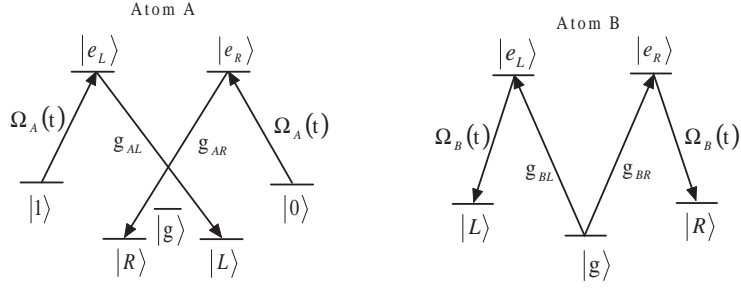}
\caption{The level configurations of atom A and B.}
\end{figure}
\section{Model and effective dynamics generation of two atoms three-dimensional entanglement}
The schematic setup for generating three-dimensional entanglement
of two atoms is shown in Fig.~1. We consider a cavity-fibre-cavity
system, in which two atoms are trapped in the corresponding
optical cavities connected by a fiber. Under the short fiber limit
$(lv)/(2\pi c)\ll 1$, only the resonant mode of the fiber will
interact with the cavity mode \cite{SMB2006}, where $ l$ is the
length of the fiber and $v$ is the decay rate of the cavity field
into a continuum of fiber modes. The corresponding level
structures of atoms are shown in Fig.~2. Atom A has two excited
states $|e_{L}\rangle,|e_{R}\rangle$, and five ground states
$|1\rangle$, $|R\rangle$, $|L\rangle$, $|g\rangle$ and
$|0\rangle$, while atom B is a five-level system with three ground
states $|R\rangle$, $|L\rangle$ and $|g\rangle$, two excited
states $|e\rangle_{L}$ and $|e\rangle_{R}$. For atom A, the
transitions $|0\rangle\leftrightarrow|e_{R}\rangle$ and
$|1\rangle\leftrightarrow|e_{L}\rangle$ are driven by classical
fields with the same Rabi frequency $\Omega_A(t)$. And
 the transitions
$|R\rangle\leftrightarrow|e_{R}\rangle$ and
$|L\rangle\leftrightarrow|e_{L}\rangle$ are resonantly driven by
the corresponding cavity mode $a_{Aj}$ with $j$-circular
polarization and the coupling strength is $g_{Aj}$ $(j=L,R)$. For
atom B, the transitions $|R\rangle\leftrightarrow|e_{R}\rangle$
and $|L\rangle\leftrightarrow|e_{L}\rangle$ are driven by
classical fields with the same Rabi frequency $\Omega_B(t)$, and
the transitions $|g\rangle\leftrightarrow|e_{R}\rangle$ and
$|g\rangle\leftrightarrow|e_{L}\rangle$ are resonantly driven by
the corresponding cavity mode $a_{Bj}$ with $j$-circular
polarization and the coupling strength is $g_{Bj}$ $(j=L,R)$. The
whole Hamiltonian in the interaction picture can be written as
($\hbar=1$):
\begin{eqnarray}\label{1}
H_{\rm total}&=&H_{a\text{-}l}+H_{a\text{-}c\text{-}f},
\end{eqnarray}
\begin{eqnarray}\label{2}
H_{a\text{-}l}&=&\Omega_{A}(t)(|e_L\rangle_{A}\langle{1}|+|e_R\rangle_{A}\langle{0}|)+\Omega_{B}(t)(|e_L\rangle_{B}\langle{L}|
+|e_R\rangle_{B}\langle{R}|)+\rm H.c.,
\end{eqnarray}
\begin{eqnarray}\label{3}
H_{a\text{-}c\text{-}f}&=&g_{AL}a_{AL}|e_L\rangle_{A}\langle{L}|+g_{AR}a_{AR}|e_R\rangle_{A}\langle{R}|
+g_{BL}a_{BL}|e_L\rangle_{B}\langle{g}|+g_{BR}a_{BR}|e_R\rangle_{B}\langle{g}|\cr
&&+\eta b_{L}(a_{AL}^{\dag}+a_{BL}^{\dag})+\eta
b_{R}(a_{AR}^{\dag}+a_{BR}^{\dag})+\rm H.c.,
\end{eqnarray}
where $\eta$ is the coupling strength between cavity mode and the
fiber mode, $b_{R(L)}$ is the annihilation operator for the fiber
mode with $R(L)$-circular polarization, $a_{A(B)R(L)}$ is the
annihilation operator for the corresponding cavity field with
$R(L)$-circular polarization, and $g_{A(B)R(L)}$ is the coupling
strength between the corresponding cavity mode and the trapped
atom.

In order to obtain the following two atoms three-dimensional
entanglement:
\begin{eqnarray}\label{4}
|\Psi\rangle&=&\frac{1}{\sqrt{3}}(|R\rangle_A|R\rangle_B+|L\rangle_A|L\rangle_B+|g\rangle_A|g\rangle_B),
\end{eqnarray}
we assume atom A in the state
$\frac{1}{\sqrt{3}}(|1\rangle_A+|0\rangle_A+|g\rangle_A)$, while
atom B in the state $|g\rangle_B$, both the cavity modes and the
fiber mode in vacuum state
$|0\rangle_{AC}|0\rangle_{BC}|0\rangle_f$, and then demonstrate
that with the help of QZD the atom state $|0\rangle_A|g\rangle_B$
can be adiabatically evolved to $|R\rangle_A|R\rangle_B$, and
$|1\rangle_A|g\rangle_B$ can be adiabatically evolved to
$|L\rangle_A|L\rangle_B$. It is easy to know that
$|g\rangle_A|g\rangle_B$ will remain unchange since there is no
excitation for all the field modes at the beginning.
 For the initial state
$|0\rangle_A|g\rangle_B|0\rangle_{AC}|0\rangle_{BC}|0\rangle_f$,
the whole system evolves in the subspace spanned by
\begin{eqnarray}\label{5}
|\phi_1\rangle&=&|0\rangle_A|g\rangle_B|0\rangle_{AC}|0\rangle_{BC}|0\rangle_f,\nonumber\\
|\phi_2\rangle&=&|e_R\rangle_A|g\rangle_B|0\rangle_{AC}|0\rangle_{BC}|0\rangle_f,\nonumber\\
|\phi_3\rangle&=&|R\rangle_A|g\rangle_B|1_R\rangle_{AC}|0\rangle_{BC}|0\rangle_f,\nonumber\\
|\phi_4\rangle&=&|R\rangle_A|g\rangle_B|0\rangle_{AC}|0\rangle_{BC}|1_R\rangle_f,\nonumber\\
|\phi_5\rangle&=&|R\rangle_A|g\rangle_B|0\rangle_{AC}|1_R\rangle_{BC}|0\rangle_f,\nonumber\\
|\phi_6\rangle&=&|R\rangle_A|e_R\rangle_B|0\rangle_{AC}|0\rangle_{BC}|0\rangle_f,\nonumber\\
|\phi_7\rangle&=&|R\rangle_A|R\rangle_B|0\rangle_{AC}|0\rangle_{BC}|0\rangle_f.\nonumber\\
\end{eqnarray}
Under the condition $\Omega_{A}(t),\Omega_{B}(t)\ll
\eta,g_{AR(L)}, g_{BR(L)}$,
 the Hilbert subspace can be divided into
five invariant Zeno subspaces \cite{PS2002,PGS2009}:
\begin{eqnarray}\label{6}
\Gamma_{P1}&=&\Big\{|\phi_1\rangle,|\phi_7\rangle,|\psi_1\rangle\Big\},\nonumber\\
\Gamma_{P2}&=&\Big\{|\psi_2\rangle\Big\},~~~~~~~~~\Gamma_{P3}~=~\Big\{|\psi_3\rangle\Big\},\nonumber\\
\Gamma_{P4}&=&\Big\{|\psi_4\rangle\Big\},~~~~~~~~~\Gamma_{P5}~=~\Big\{|\psi_5\rangle\Big\},
\end{eqnarray}
with the eigenvalues $\lambda_1=0$, $\lambda_2=-g$, $\lambda_3=g$,
$\lambda_4=-\sqrt{g^2+2\eta^2}=-\epsilon$, and
$\lambda_5=\sqrt{g^2+2\eta^2}=\epsilon$, where we assume
$g_{AR(L)}=g_{BR(L)}=g$ for simplicity. Here
\begin{eqnarray}\label{7}
|\psi_1\rangle&=&\frac{1}{\epsilon}(\eta|\phi_2\rangle-g|\phi_4\rangle+\eta|\phi_6\rangle),\nonumber\\
|\psi_2\rangle&=&\frac{1}{2}(-|\phi_2\rangle+|\phi_3\rangle-|\phi_5\rangle+|\phi_6\rangle),\nonumber\\
|\psi_3\rangle&=&\frac{1}{2}(-|\phi_2\rangle-|\phi_3\rangle+|\phi_5\rangle+|\phi_6\rangle),\nonumber\\
|\psi_4\rangle&=&\frac{1}{2\epsilon}(g|\phi_2\rangle-\epsilon|\phi_3\rangle+2\eta|\phi_4\rangle)-\epsilon|\phi_5\rangle+g|\phi_6\rangle,\nonumber\\
|\psi_5\rangle&=&\frac{1}{2\epsilon}(g|\phi_2\rangle+\epsilon|\phi_3\rangle+2\eta|\phi_4\rangle)+\epsilon|\phi_5\rangle+g|\phi_6\rangle,
\end{eqnarray}
and the corresponding projection
\begin{eqnarray}\label{8}
P_i^\alpha =
|\alpha\rangle\left\langle\alpha\right|,(|\alpha\rangle\in\Gamma_{Pi}).
\end{eqnarray}
Under the above condition, the system Hamiltonian can be rewritten
as the following form \cite{PGS2009}:
\begin{eqnarray}\label{9}
H_{\rm total}&\simeq&
\sum_{i,\alpha,\beta}(\lambda_iP_i^\alpha+P_i^\alpha H_{a\text{-}l} P_i^\beta) \nonumber\\
&=&-g|\psi_2\rangle\left\langle\psi_2\right|+g|\psi_3\rangle\left\langle\psi_3\right|-\epsilon|\psi_4\rangle\left\langle\psi_4\right|+
\epsilon|\psi_5\rangle\left\langle\psi_5\right|\cr
&&+\frac{1}{\epsilon}\eta(\Omega_{A}(t)|\psi_1\rangle\left\langle\phi_1\right|+\Omega_{B}(t)|\psi_1\rangle\left\langle\phi_7\right|+\rm
H.c.).
\end{eqnarray}
When we choose the initial state $|\phi_1\rangle =
|0\rangle_A|g\rangle_B|0\rangle_{AC}|0\rangle_{BC}|0\rangle_f$,
 the Hamiltonian $H_{\rm total}$ reduces to
 \begin{eqnarray}\label{10}
H_{\rm
eff}&=&\Omega_{A1}(t)|\psi_1\rangle\left\langle\phi_1\right|+\Omega_{B1}(t)|\psi_1\rangle\left\langle\phi_7\right|+\rm
H.c.,
\end{eqnarray}
where $\Omega_{A1}(t)=\frac{1}{\epsilon}\eta\Omega_{A}(t)$ and
$\Omega_{B1}(t)=\frac{1}{\epsilon}\eta\Omega_{B}(t)$. Combining
with adiabatic passage method, we can obtain the dark state of
$H_{\rm eff}$:
\begin{eqnarray}\label{11}
|\psi_{D1}\rangle&=&\frac{1}{\sqrt{\Omega_{A1}(t)^2+\Omega_{B1}(t)^2}}(-\Omega_{B1}(t)|\phi_1\rangle+\Omega_{A1}(t)|\phi_7\rangle).
\end{eqnarray}
 When the pulses shape satisfy
\begin{eqnarray}\label{12}
\lim_{t\to-\infty}\frac{\Omega_{A1}(t)}{\Omega_{B1}(t)}=0,~
\lim_{t\to+\infty}\frac{\Omega_{A1}(t)}{\Omega_{B1}(t)}=\infty,
\end{eqnarray}
thus, based on the effective Hamiltonian (\ref{10}), our proposal
for population transfer from $|\phi_1\rangle$ to $|\phi_7\rangle$
can be achieved.

On the other hand, if the initial state is
$|1\rangle_A|g\rangle_B|0\rangle_{AC}|0\rangle_{BC}|0\rangle_f$,
the whole system evolves in the subspace spanned by
\begin{eqnarray}\label{13}
|\phi_1'\rangle&=&|1\rangle_A|g\rangle_B|0\rangle_{AC}|0\rangle_{BC}|0\rangle_f,\nonumber\\
|\phi_2'\rangle&=&|e_L\rangle_A|g\rangle_B|0\rangle_{AC}|0\rangle_{BC}|0\rangle_f,\nonumber\\
|\phi_3'\rangle&=&|L\rangle_A|g\rangle_B|1_L\rangle_{AC}|0\rangle_{BC}|0\rangle_f,\nonumber\\
|\phi_4'\rangle&=&|L\rangle_A|g\rangle_B|0\rangle_{AC}|0\rangle_{BC}|1_L\rangle_f,\nonumber\\
|\phi_5'\rangle&=&|L\rangle_A|g\rangle_B|0\rangle_{AC}|1_L\rangle_{BC}|0\rangle_f,\nonumber\\
|\phi_6'\rangle&=&|L\rangle_A|e_L\rangle_B|0\rangle_{AC}|0\rangle_{BC}|0\rangle_f,\nonumber\\
|\phi_7'\rangle&=&|L\rangle_A|L\rangle_B|0\rangle_{AC}|0\rangle_{BC}|0\rangle_f.
\end{eqnarray}
In this situation, with the method mentioned above, we can easily
obtain the effective Hamiltonian:
\begin{eqnarray}\label{14}
H_{\rm
eff}'&=&\Omega_{A1}(t)|\psi_1'\rangle\left\langle\phi_1'\right|+\Omega_{B1}(t)|\psi_1'\rangle\left\langle\phi_7'
 \right|+\rm H.c.,
\end{eqnarray}
where
$|\psi_1'\rangle=\frac{1}{\epsilon}(\eta|\phi_2'\rangle-g|\phi_4'\rangle+\eta|\phi_6'\rangle)$,
$\Omega_{A1}(t)=\frac{1}{\epsilon}\eta\Omega_{A}(t)$ and
$\Omega_{B1}(t)=\frac{1}{\epsilon}\eta\Omega_{B}(t)$. Combining
with adiabatic passage method, we can obtain the dark state of
$H_{\rm eff}'$:
\begin{eqnarray}\label{15}
|\psi_{D2}\rangle&=&\frac{1}{\sqrt{\Omega_{A1}(t)^2+\Omega_{B1}(t)^2}}(-\Omega_{B1}(t)|\phi_1'\rangle+\Omega_{A1}(t)|\phi_7'\rangle).
\end{eqnarray}
When the pulses shape satisfy Eq.(\ref{12}), the initial state
$|\phi_1'\rangle$ involves to $|\phi_7'\rangle$ eventually.

If the initial state is
$|g\rangle_A|g\rangle_B|0\rangle_{AC}|0\rangle_{BC}|0\rangle_f$,
it will not change at all during the whole evolution.

Therefore, the initial state
\begin{eqnarray}\label{16}
\Psi(0)=\frac{1}{\sqrt{3}}(|0\rangle_A+|1\rangle_A+|g\rangle_A)|g\rangle_B|0\rangle_{AC}|0\rangle_{BC}|0\rangle_f
\end{eqnarray}
of the compound system will evolve to the state
\begin{eqnarray}\label{17}
\Psi(t)=\frac{1}{\sqrt{3}}(|R\rangle_A|R\rangle_B+|L\rangle_A|L\rangle_B+|g\rangle_A|g\rangle_B)|0\rangle_{AC}|0\rangle_{BC}|0\rangle_f,
\end{eqnarray}
which is a product state of the two atoms three-dimensional
entanglement, the cavity modes vacuum state, and the fiber mode
vacuum state.
\begin{figure}[htb]\centering
\includegraphics[width=10cm]{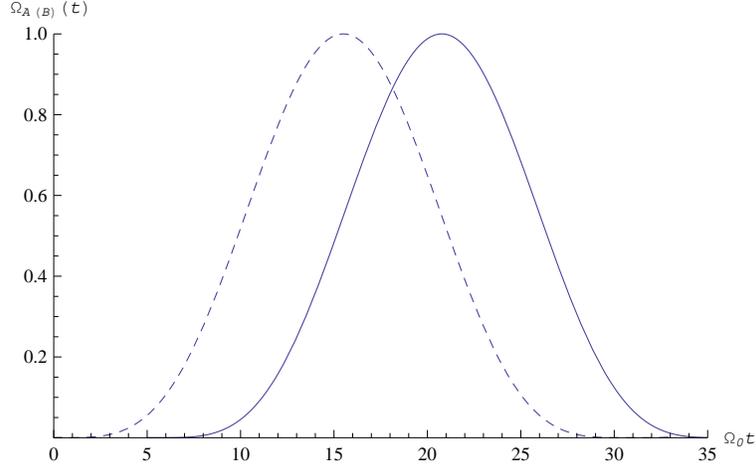}
\caption{The time dependence of the laser fields $\Omega_A(t)$
corresponding to solid line and $\Omega_B(t)$ corresponding to
dashed line with $t_0=\Omega_0^{-1}$ and $\tau=5.27 t_0$.}
\end{figure}
\begin{figure}[htb]\centering
\includegraphics[width=10cm]{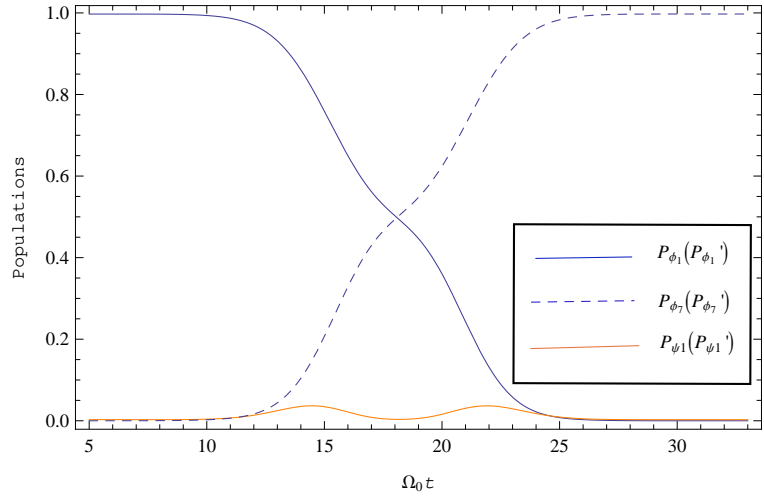}
\caption{Time evolutions of the populations corresponding to the
system states versus $\Omega_0t$ with $g=20\Omega_0 $,
$\eta=100g$, $t_0=\Omega_0^{-1} $ and $\tau=5.27 t_0$.}
\end{figure}

\section{Numerical analysis and the robustness of the scheme}

In order to generate two atoms three-dimensional entanglement, the
conditions of Eq.~(\ref{12}) should be satisfied in our scheme.
For this reason, we can choose the pulses shape of the laser
fields $\Omega_A(t)$ and $\Omega_B(t)$ in the original Hamiltonian
$H_{\rm total}$ as:
\begin{eqnarray}\label{18}
\Omega_{A}(t)= \Omega_{0}\sin^4{[\pi(t-\tau)/{31 t_0}]}
\end{eqnarray}
and
\begin{eqnarray}\label{19}
\Omega_{B}(t)= \Omega_{0}\sin^4{(\pi t/{31 t_0})}.
\end{eqnarray}
Here $\Omega_{0}$ is the pulse amplitude, $\tau$ being the time
delay. Fig.~3 shows the Rabi frequencies $\Omega_{A}(t)$ and
$\Omega_{B}(t) $ versus $\Omega_0 t $ with $t_0=\Omega_0^{-1}$ and
$\tau=5.27 t_0$. The Rabi frequencies are two delayed but
partially overlapped pulses. The population curves of
$|\phi_1\rangle(|\phi_1'\rangle)$,
$|\phi_7\rangle(|\phi_7'\rangle)$ and
$|\psi_1\rangle(|\psi_1'\rangle)$ versus $\Omega_0 t$ are depicted
in Fig.~4, where we choose $g=20\Omega_0 $, $\eta=100g$,
$t_0=\Omega_0^{-1} $ and $\tau=5.27 t_0$. From Fig.~4 we can see
that the population inverts completely when $\Omega_{0}t$ is over
25.~Through the above processes, we can generate two atoms
three-dimensional entanglement successfully. The evolutions are
governed by the effectively Hamiltonian $H_{\rm eff}(H_{\rm
eff}')$, and $\Omega_A(t)$ and $\Omega_B(t)$ are defined by
Eqs.(\ref{18}) and (\ref{19}), respectively.
\begin{figure}[htb]\centering
\includegraphics[width=10cm]{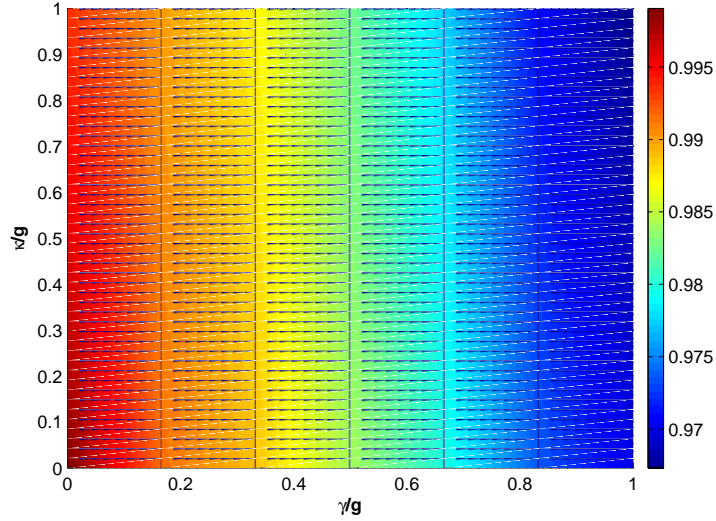}
\caption{The fidelity corresponding to the target state versus
$\kappa/g$ and $\gamma/g$, with $g=20\Omega_0 $, $\eta=100g$,
$t_0=\Omega_0^{-1} $ and $\tau=5.27 t_0$.}
\end{figure}

It is well-known that whether a scheme is applicable for quantum
information processing and quantum computing depends on the
robustness against possible mechanisms of decoherence. To examine
the robustness of our scheme described in the previous sections,
we consider the effect of photon leakage and atom spontaneous
decay. The corresponding master equation for the whole system
density matrix $\rho(t)$ has the following form:
\begin{eqnarray}\label{20}
\dot{\rho(t)}&=&-i[H_{\rm
total},\rho(t)]-\sum_{j=L,R}\frac{\kappa_f}{2}[b_j^+b_j\rho(t)-2b_j\rho(t)b_j^++\rho(t)b_j^+b_j]\cr
&&-\sum_{j=L,R}\sum_{i=A,B}\frac{\kappa_j}{2}[a_{ij}^+a_{ij}\rho(t)-2a_{ij}\rho(t)a_{ij}^++\rho(t)a_{ij}^+a_{ij}]\cr
&&-\frac{\gamma_A}{2}\{\sum_{h=L,1}[\sigma_{e_L,e_L}^A\rho(t)-2\sigma_{h,e_L}^A\rho(t)\sigma_{e_L,h}^A+\rho(t)\sigma_{e_L,e_L}^A]\cr
&&+\sum_{k=R,0}[\sigma_{e_R,e_R}^A\rho(t)-2\sigma_{k,e_R}^A\rho(t)\sigma_{e_R,k}^A+\rho(t)\sigma_{e_R,e_R}^A]\}\cr
&&-\sum_{j=L,R}\sum_{l=j,g}\frac{\gamma_B}{2}[\sigma_{e_j,e_j}^B\rho(t)-2\sigma_{l,e_j}^B\rho(t)\sigma_{e_j,l}^B+\rho(t)\sigma_{e_j}^B\sigma_{e_j}^B],
\end{eqnarray}
where $H_{\rm total}$ is given by Eq.~(1). $\kappa_f$ and
$\kappa_{R(L)}$ are the photon leakage rates of the fiber mode and
cavity mode $R(L)$, respectively. $\gamma_{A(B)}$ is the atom A(B)
spontaneous emission rate from the excited state
$|e_{R}\rangle(|e_{L}\rangle)$ to the ground state
$|R\rangle(|L\rangle)$ and $|g\rangle$, respectively.
$\sigma_{m,n}=|m\rangle\left\langle
n\right|(m,n=0,1,L,R,g,e_L,e_R)$. For simplicity, we assume
$\kappa_f=\kappa_R=\kappa_L=\kappa/2$,
$\gamma_A=\gamma_B=\gamma/4$ and the initial condition $\rho(0)=
|\Psi_0\rangle\left\langle \Psi_0\right|$. In Fig.~5, the fidelity
of the final two atoms three-dimensional entanglement is plotted
versus the dimensionless parameters $\kappa/g$ and $\gamma/g$ by
numerically solving the master Eq.~(\ref{20}). From Fig.~5 we can
see that the fidelity of two atoms three-dimensional entanglement
is higher than $96.5\%$ even in the range of $\kappa $ and $\gamma
$ close to $g$.

\section{Generation of $N$-atom three-dimensional
entanglement}\label{section4}

\begin{figure}[htb]\centering
\includegraphics[width=10cm]{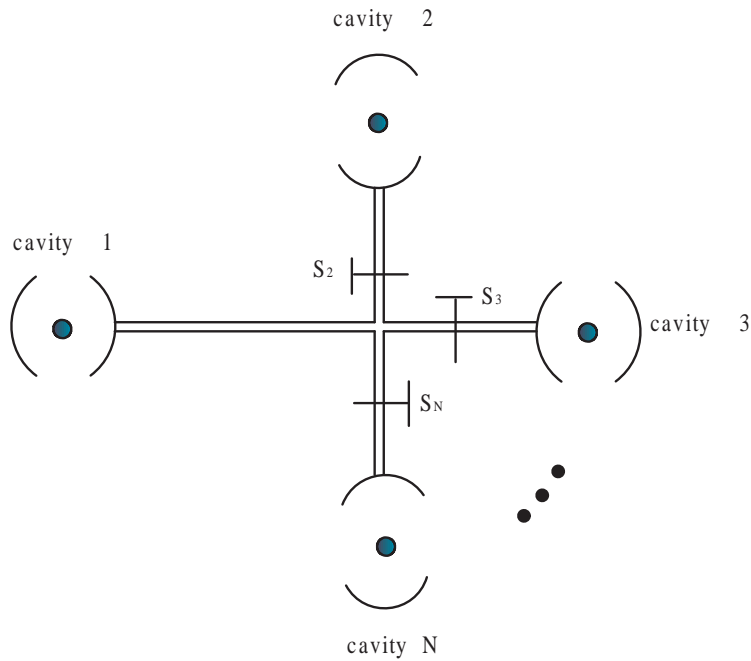}
\caption{Schematic setup for generation of $N$-atom
three-dimensional entanglement.}
\end{figure}

In this section, we will show how to deterministically generate
the $N$-atom three-dimensional entanglement.~The schematic setup
is shown in Fig.~6, where $N$ atoms are trapped in $N$ spatially
separated cavities connected by optical fibers. $S_2, S_3,...,S_N$
are the optical switch devices which are used to control the
interaction between the $n\text{-}$th cavity and the first cavity.
The atom in cavity 1 possesses the level structure as atom A
presented in Fig.~2, and the atoms in the other $(n=2,3,...,N)$
cavities are the same as atom B given in Fig.~2. By only turning
on the $n\text{-}$th switch S$_n$ ($n$ = 2,3,..., $N$), and the
others off, the interaction between the $1st$ atom and the
$n\text{-}$th atom can be achieved. In this case, the interaction
Hamiltonian has the following form ($\hbar=1$):
\begin{eqnarray}\label{21}
H_{\rm
total}^{n}&=&H_{a\text{-}l}^{n}+H_{a\text{-}c\text{-}f}^{n},
\end{eqnarray}
\begin{eqnarray}\label{22}
H_{a\text{-}l}^{n}&=&\Omega_{A}(t)(|e_L\rangle_{1}\langle{1}|+|e_R\rangle_{1}\langle{0}|)+\Omega_{B}(t)(|e_L\rangle_{n}\langle{L}|
+|e_R\rangle_{n}\langle{R}|)+\rm H.c.,
\end{eqnarray}
\begin{eqnarray}\label{23}
H_{a\text{-}c\text{-}f}^{n}&=&g_{1L}a_{1L}|e_L\rangle_{1}\langle{L}|+g_{1R}a_{1R}|e_R\rangle_{1}\langle{R}|
+g_{nL}a_{nL}|e_L\rangle_{n}\langle{g}|\cr
&&+g_{nR}a_{nR}|e_R\rangle_{n}\langle{g}|+\eta
b_{L}(a_{1L}^{\dag}+a_{nL}^{\dag})+\eta
b_{R}(a_{1R}^{\dag}+a_{nR}^{\dag})+\rm H.c.,
\end{eqnarray}
where $``n"$ represents the atom trapped in the $n\text{-}$th
cavity, $a_{nR(L)}$ is annihilation operator corresponding to the
$n\text{-}$th cavity with $R(L)$-circular polarization, and
$g_{nR(L)}$ is coupling strength between the corresponding cavity
mode and the trapped atom.

In order to obtain ~$N$-atom three-dimensional entanglement
\begin{eqnarray}\label{24}
|\varphi\rangle_N&=&\frac{1}{\sqrt{3}}(|R\rangle_1|R\rangle_2....|R\rangle_N
+|L\rangle_1|L\rangle_2....|L\rangle_N+|g\rangle_1|g\rangle_2....|g\rangle_N),
\end{eqnarray}
we assume $g_{nR}=g_{nL}=g$ for simplicity, and the initial state
of the compound system is:
\begin{eqnarray}\label{25}
|\varphi\rangle_1&=&\frac{1}{\sqrt{3}}(|0\rangle_1+|1\rangle_1
+|g\rangle_1)|g\rangle_2|g\rangle_3....|g\rangle_N|0\rangle_{1C}|0\rangle_{2C}....|0\rangle_{NC}|0\rangle_{f}.
\end{eqnarray}

(i) First, turn on S$_2$ and keep other switches off, then the
first two atoms can be prepared in the state $|\Psi\rangle =
\frac{1}{\sqrt{3}}(|R\rangle_1|R\rangle_2+|L\rangle_1|L\rangle_2+|g\rangle_1|g\rangle_2)$
with the method mentioned in Sec. II.

(ii) Then, turn off S$_2$, and apply another pulse on atom 1 to
drive the transitions $|L\rangle_1\rightarrow|1\rangle_1$ and
$|R\rangle_1\rightarrow|0\rangle_1$, which leads the compound
system to the state:
\begin{eqnarray}\label{26}
|\varphi\rangle_2&=&\frac{1}{\sqrt{3}}(|0\rangle_1|R\rangle_2+|1\rangle_1|L\rangle_2
+|g\rangle_1|g\rangle_2)|g\rangle_3|g\rangle_4....|g\rangle_N|0\rangle_{1C}|0\rangle_{2C}....|0\rangle_{NC}|0\rangle_{f}.
\end{eqnarray}

(iii) Subsequently, only S$_n$ $(n = 3,..., N)$ is turned on, thus
the $1st$ atom can interact with the $n\text{-}$th atom. Next,
perform the same operations as the first step successively to make
the state in Eq. (\ref{26}) evolve to
\begin{eqnarray}\label{27}
|\varphi\rangle_3&=&\frac{1}{\sqrt{3}}(|R\rangle_1|R\rangle_2...|R\rangle_n+|L\rangle_1|L\rangle_2....|L\rangle_n
+|g\rangle_1|g\rangle_2....|g\rangle_n)|g\rangle_{n+1}|g\rangle_{n+2}\cr
&&....|g\rangle_N|0\rangle_{1C}|0\rangle_{2C}....|0\rangle_{NC}|0\rangle_{f}.
\end{eqnarray}

(iv) Finally we can obtain the $N$-atom three-dimensional
entanglement after repeating the steps (ii) and (iii) $(N-2)$
times. Thus, the state of the compound system finally evolves to

\begin{eqnarray}\label{28}
|\varphi\rangle_4&=&\frac{1}{\sqrt{3}}(|R\rangle_1|R\rangle_2....|R\rangle_N
+|L\rangle_1|L\rangle_2....|L\rangle_N\cr
&&+|g\rangle_1|g\rangle_2....|g\rangle_N)|0\rangle_{1C}|0\rangle_{2C}....|0\rangle_{NC}|0\rangle_{f}.
\end{eqnarray}
That is a product state of $N$-atom three-dimensional
entanglement, the cavity modes state, and the fiber modes state.

\section{Generation of high-dimensional
entanglement of two spatially separated atoms}\label{section5}

\begin{figure}[htb]\centering
\includegraphics[width=10cm]{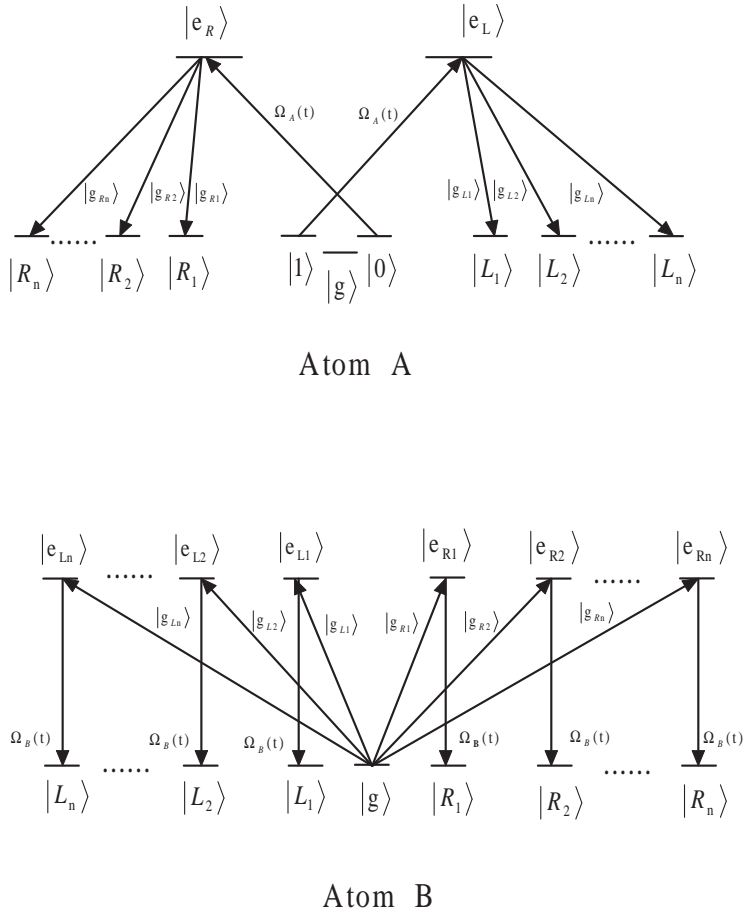}
\caption{The potential atomic levels configuration for atom A and
B.}
\end{figure}

We note that the scheme can also be expanded to generate
high-dimensional entanglement of two spatially separated atoms.
The potential atomic configurations is plotted in Fig.~7. We
assume that the cavity supports $2N$ independent modes of photon
fields.~Then the Hamiltonian can be written as ($\hbar=1$):
\begin{eqnarray}\label{29}
\mathcal{H}_{\rm
total}&=&\mathcal{H}_{a\text{-}l}+\mathcal{H}_{a\text{-}c\text{-}f},
\end{eqnarray}
\begin{eqnarray}\label{30}
\mathcal{H}_{a\text{-}l}&=&\Omega_{A}(t)(|e_L\rangle_{A}\langle{1}|+|e_R\rangle_{A}\langle{0}|)+\Omega_{B}(t)\sum_{i=1}^N(|e_{Li}\rangle_{B}\langle{L_i}|
+|e_{Ri}\rangle_{B}\langle{R_i}|)+\rm H.c.,
\end{eqnarray}
\begin{eqnarray}\label{31}
\mathcal{H}_{a\text{-}c\text{-}f}&=&\sum_{i=1}^N[(g_{ALi}a_{ALi}|e_L\rangle_{A}\langle{L_i}|+g_{ARi}a_{ARi}|e_R\rangle_{A}\langle{R_i}|)
+g_{BLi}a_{BLi}|e_{Li}\rangle_{B}\langle{g}|\cr
&&+g_{BRi}a_{BRi}|e_{Ri}\rangle_{B}\langle{g}| +\eta
b_{Li}(a_{ALi}^{\dag}+a_{BLi}^{\dag})+\eta
b_{Ri}(a_{ARi}^{\dag}+a_{BRi}^{\dag})+\rm H.c.].
\end{eqnarray}
Assume the initial state of the whole system is
$|1\rangle_A|g\rangle_B|0_10_2....0_N\rangle_{AC}$
$|0_10_2....0_N\rangle_{BC}|0_10_2....0_N\rangle_f$, then the
system will evolve in the subsystem $\Gamma$ spanned by the
vectors
$\{|\zeta_1\rangle,|\zeta_2\rangle....|\zeta_{5N+2}\rangle\}$:
\begin{eqnarray}\label{32}
\Gamma=\{|\zeta_1\rangle&=&|1\rangle_A|g\rangle_B|0_10_2....0_N\rangle_{AC}|0_10_2....0_N\rangle_{BC}|0_10_2....0_N\rangle_f,\nonumber\\
|\zeta_2\rangle&=&|e_L\rangle_A|g\rangle_B|0_10_2....0_N\rangle_{AC}|0_10_2....0_N\rangle_{BC}|0_10_2....0_N\rangle_f,\nonumber\\
|\zeta_{2+i}\rangle&=&|L_i\rangle_A|g\rangle_B|....1i....\rangle_{AC}|0_10_2....0_N\rangle_{BC}|0_10_2....0_N\rangle_f,\nonumber\\
|\zeta_{2+N+i}\rangle&=&|L_i\rangle_A|g\rangle_B|0_10_2....0_N\rangle_{AC}|0_10_2....0_N\rangle_{BC}|....1i....\rangle_f,\nonumber\\
|\zeta_{2+2N+i}\rangle&=&|L_i\rangle_A|g\rangle_B|0_10_2....0_N\rangle_{AC}|....1i....\rangle_{BC}|0_10_2....0_N\rangle_f,\nonumber\\
|\zeta_{2+3N+i}\rangle&=&|L_i\rangle_A|e_{Li}\rangle_B|0_10_2....0_N\rangle_{AC}|0_10_2....0_N\rangle_{BC}|0_10_2....0_N\rangle_f,\nonumber\\
|\zeta_{2+4N+i}\rangle&=&|L_i\rangle_A|L_i\rangle_B|0_10_2....0_N\rangle_{AC}|0_10_2....0_N\rangle_{BC}|0_10_2....0_N\rangle_f.\},
\end{eqnarray}
where $i =1,2,3,...,N$. With the prior procedures presented in
Sec. II, we can obtain the effective Hamiltonian:
 \begin{eqnarray}\label{33}
 \mathcal{H}_{\rm
eff}&=&\frac{1}{\sqrt{N+1}}(\Omega_{A1}(t)|X_1\rangle\left\langle\zeta_1\right|
+\Omega_{B1}(t)\sum_{j=4N+3}^{5N+2}|X_1\rangle\left\langle\zeta_j\right|+\rm
H.c.),
\end{eqnarray}
with $\Omega_{A1}(t)=\frac{1}{\epsilon}\eta\Omega_{A}(t)$ and
$\Omega_{B1}(t)=\frac{1}{\epsilon}\eta\Omega_{B}(t)$, where
\begin{eqnarray}\label{34}
|X_1\rangle=\frac{1}{\epsilon\sqrt{2N+1}}(\eta|\zeta_2\rangle-g\sum_{j=N+3}^{2N+2}|\zeta_j\rangle+\eta\sum_{k=3N+3}^{4N+2}|\zeta_k\rangle).
\end{eqnarray}
With the help of adiabatic passage method, we can obtain the dark
state of $\mathcal{H}_{\rm eff}$:
\begin{eqnarray}\label{35}
|X_{D}\rangle&=&\frac{1}{\sqrt{(N+1)(\Omega_{A1}(t)^2+\Omega_{B1}(t)^2)}}(-\Omega_{B1}(t)|\zeta_1\rangle+\Omega_{A1}(t)\sum_{j=4N+3}^{5N+2}|\zeta_j\rangle).
\end{eqnarray}
When Eq.~(\ref{12}) is satisfied, the initial state
$|1\rangle_A|g\rangle_B|0_10_2....0_N\rangle_{AC}
|0_10_2....0_N\rangle_{BC}$ $|0_10_2....0_N\rangle_f$ of the whole
system will finally evolve to
\begin{eqnarray}\label{36}
|\nu\rangle&=&
\frac{1}{\sqrt{N}}\sum_{j=4N+3}^{5N+2}|\zeta_j\rangle\cr
&=&\frac{1}{\sqrt{N}}\sum_{m=1}^{N}|L_m\rangle_A|L_m\rangle_B|0_10_2....0_N\rangle_{AC}|0_10_2....0_N\rangle_{BC}|0_10_2....0_N\rangle_f,
\end{eqnarray}
which is a $N$-dimensional maximally entanglement.

For the initial state
$|0\rangle_A|g\rangle_B|0_10_2....0_N\rangle_{AC}$
$|0_10_2....0_N\rangle_{BC}|0_10_2....0_N\rangle_f$, by using the
similar way from Eqs.~(\ref{29})-(\ref{36}), this initial state
finally evolves to
\begin{eqnarray}\label{37}
|\nu_1\rangle&=&\frac{1}{\sqrt{N}}\sum_{k=1}^{N}|R_k\rangle_A|R_k\rangle_B|0_10_2....0_N\rangle_{AC}|0_10_2....0_N\rangle_{BC}|0_10_2....0_N\rangle_f.
\end{eqnarray}
On the other hand, the initial state
$|g\rangle_A|g\rangle_B|0_10_2....0_N\rangle_{AC}$$|0_10_2....0_N\rangle_{BC}|0_10_2....0_N\rangle_f$
don't participate in the evolution.

Therefore, if we choose the initial state of the combined system
as
\begin{eqnarray}\label{38}
|\nu\rangle_0=\frac{1}{\sqrt{3}}(|1\rangle_A+|0\rangle_A+|g\rangle_A)|g\rangle_B|0_10_2....0_N\rangle_{AC}|0_10_2....0_N\rangle_{BC}|0_10_2....0_N\rangle_f,
\end{eqnarray}
after implementing all the operations mentioned above, the
high-dimensional entanglement can be obtained as the following
form:
\begin{eqnarray}\label{39}
|\beta\rangle&=&\frac{1}{\sqrt{2N+1}}[\sum_{m=1}^{N}|L_m\rangle_A|L_m\rangle_B+\sum_{k=1}^{N}|R_k\rangle_A|R_k\rangle_B\cr
&&+|g\rangle_A|g\rangle_B]|0_10_2....0_N\rangle_{AC}|0_10_2....0_N\rangle_{BC}|0_10_2....0_N\rangle_f.
\end{eqnarray}

\section{Analysis and discussion}

We now analyze the feasibility of the experiment for this scheme.
The appropriate atomic level configuration can be obtained from
the hyperfine structure of ${}^{133}\!$Cs. 5S$_{1/2}$ ground level
$|F=3,m=2\rangle(|F=3,m=-2\rangle)$ corresponds to
$|R\rangle(|L\rangle)$ and $|F=2,m=1\rangle(|F=2,m=-1\rangle)$
corresponds to $|0\rangle(|1\rangle)$, respectively, while
5P$_{3/2}$ excited level $|F=3,m=1\rangle(|F=3,m=-1\rangle)$
corresponds to $|e_R\rangle(|e_L\rangle)$. Other hyperfine levels
in the ground-state manifold can be used as $|g\rangle$ for atom
A. For atom B, the states $|R\rangle,|L\rangle$ and $|g\rangle$
correspond to $|F=2,m=-1\rangle,|F=2,m=1\rangle$ and
$|F=3,m=0\rangle$ of 5S$_{1/2}$ ground levels, respectively. And
$|e_R\rangle(|e_L\rangle)$ corresponds to
$|F=3,m=-1\rangle(|F=3,m=1\rangle)$ of 5P$_{3/2}$ excited level.
In this paper, we choose $\Omega_{A(B)}(t)/g$ is less than 0.05,
so that the Zeno condition can be satisfied well. In experiments,
the cavity QED parameters $g=2.5GHz$, $\kappa=10MHz$, and
$\gamma=10MHz$ have been realized in \cite{SMT2005,MJF2006}. For
such parameters, the fidelity of our scheme is larger than
$99.0\%$, so our scheme is robust against both the cavity decay,
the fiber loss and the atomic spontaneous radiation and may be
very promising within current experiment technology.

In summary, we have proposed a promising scheme to generate
three-dimensional entanglement with the help of QZD in the
cavity-fiber-cavity system. Because the atoms are resonant
interaction, so the speed of producing entanglement is very fast
compared with the dispersive protocols \cite{XKW2003,XZYW2005}.
Meanwhile, the influence of various decoherence processes such as
spontaneous emission and photon loss on the generation of
entanglement is also investigated. Because during the whole
process the system keeps in a Zeno subspace without exciting the
cavity field and the fiber, and all the atoms are in the ground
states, thus the scheme is robust against the cavity, fiber and
atomic decay. Numerical results show the generation of
entanglement can be achieved with a high fidelity. Besides, the
scheme can be generalized to generate $N$-atom three-dimensional
entanglement and high-dimensional entanglement.~We hope our work
will have a crucial role in promoting quantum information
processing including implementing quantum gates, performing atomic
state transfer, generating entanglement,
etc.\\

\begin{center}$\mathbf{ACKNOWLEDGMENTS}$\end{center}
This work was supported by the National Natural Science Foundation
of China under Grant Nos. 11064016 and 61068001.

\end{document}